\documentclass[journal]{IEEEtran}

\usepackage{amsmath}
\usepackage{graphicx}  
\usepackage{subfigure}  

\ifCLASSINFOpdf
\else
\fi
\usepackage{hyperref} 

\begin{document}
%
\title{Assessing FIFO and Round Robin Scheduling: Effects on Data Pipeline Performance and Energy Usage}
%
%
%
\author{
    Malobika~Roy~Choudhury\thanks{Malobika Roy Choudhury is a researcher with a Master's degree from Arizona State University, AZ 85281 USA. Email: \texttt{malobika.roychoudhury@gmail.com}}
    \and
    Akshat~Mehrotra\thanks{Akshat Mehrotra is an engineering manager pursuing a Master's degree from Harrisburg University,
    PA 039483 USA. Email: \texttt{akshat08121989@gmail.com}}
}

%
%

\markboth{Journal of \LaTeX\ 2024 }%
{Comparing Different Process Scheduling Algorithms While Evaluating Data-Pipeline Performance and Its Impact on Energy Consumption}

%



\maketitle

\begin{abstract}
In the case of compute-intensive machine learning, efficient operating system scheduling is crucial for performance and energy efficiency. This paper conducts a comparative study over FIFO and RR scheduling policies with the application of real-time machine learning training processes and data pipelines on Ubuntu-based systems. Knowing a few patterns of CPU usage and energy consumption, we identify which policy (the exclusive or the shared) provides higher performance and/or lower energy consumption for typical modern workloads. Results of this study would help in providing better operating system schedulers for modern systems like Ubuntu, working to improve performance and reducing energy consumption in compute-intensive workloads.
\end{abstract}


%
\IEEEpeerreviewmaketitle

\section{Introduction}
\IEEEPARstart{T}{he} rapid advancement of machine learning (ML) and data-intensive applications has placed unprecedented demands on computing systems. Efficient scheduling policies in operating systems are crucial for optimizing performance and energy consumption, especially when handling compute-intensive tasks. Traditional scheduling algorithms like First-In-First-Out (FIFO) and Round Robin (RR) have been widely studied for their performance and energy efficiency trade-offs. However, most existing research has not extensively evaluated these policies in the context of modern compute-intensive ML algorithms on contemporary systems.
Over the past two decades, significant innovations have transformed computing architectures, leading to differences in how scheduling policies impact system performance and energy usage compared to legacy systems. Current research necessitates a comprehensive analysis of these advancements to understand how traditional scheduling policies perform under the workloads of today's compute-intensive applications.
This study conducts an in-depth comparison of FIFO and Round Robin scheduling policies on Ubuntu systems, utilizing real-time processes, including compute-intensive ML training tasks and data pipelines. The research aims to explore the CPU usage patterns and energy consumption effects of these scheduling policies over time. By analyzing their performance with modern workloads, the study seeks to determine which scheduling policy is more suitable for compute-intensive tasks in terms of both performance and energy efficiency.
The findings of this paper will provide valuable insights into the optimization of scheduling policies for contemporary computing systems like Ubuntu. This can guide the development of more efficient operating system schedulers that help the demands of compute expensive clusters using Linux as backend in turn contributing to enhanced performance and lower energy consumption.

%
%
%
%



\section{Related Work}
Throughout the years, systems have employed several scheduling concepts. Notable scheduling methods include First Come First Serve  \cite{stallings2018operating}, Round Robin, priority scheduling, and Shortest job first\cite{tanenbaum2015modern}.Next, we look more into that of centralized scheduling policies versus distributed scheduling policies. Traditional centralized scheduling solutions predominantly depend on schedulers such as Google Borg\cite{verma2015large} and Hadoop YARN\cite{vavilapalli2013apache}. In the context of large-scale jobs, these schedulers exhibit significant latencies and distribute resources worldwide. Despite their ability to address some constraints of centralized schedulers, distributed schedulers such as Sparrow\cite{ousterhout2013sparrow} ultimately lack the comprehensive understanding of the entire cluster.
The objective of hybrid schedulers is to integrate the advantages of both centralized and decentralized systems. Their ability to alternate between centralized decision making and decentralized scheduling is contingent upon the requirements. Notable hybrid schedulers include Mercury \cite{delgado2015mercury}, Hawk\cite{chen2018hawk}, and Eagle\cite{delgado2016eagle}. Deadline-Aware scheduling\cite{haidri2020cost}, heterogeneous cluster scheduling, and energy aware scheduling are examples of advanced scheduling algorithms. Paragon\cite{delimitrou2013paragon} and Quasar\cite{delimitrou2014quasar} leverage heterogeneous cluster scheduling to assign tasks to the most appropriate hardware. AlSched\cite{zhao2015alsched} is a method designed to optimize the efficiency of CPU and GPU resource use. Efficient scheduling can be seen as the future of high-performance computing systems. A fundamental obstacle in different scheduling methods is the effective management of the compromise between performance and energy efficiency. This work examines the conventional scheduling methods and their energy consumption effects on an Ubuntu system.
Merkel A (2006) \cite{merkel2006balancing} have discussed performance vs energy trade-offs. They have proposed their work the energy aware scheduling policies\cite{akgun2014energy}  across multiprocessor systems. They conclude that dynamically distributing the workloads based on energy profiles can lead to more efficient power usage. Dhiman and Rosing (2007) \cite{dhiman2007performance}further examined this trade-off, suggesting that scheduling strategies often optimize for either performance or energy efficiency but not both simultaneously. Arwa Zakrya et all \cite{zakrya2020comparative} (2020) concluded that partitioned scheduling algorithms are better in performance than global scheduling algorithms in some parameters which are makespan, waiting time, missed deadlines and task preemptions.  Aftab Ahmed Chandioa. Et all (2019)\cite{chandio2019comparative} concluded that there is not one scheduling policy that can satisfy the energy needs and resource management in a VM.They highlighted the necessity for a dynamic and adaptive strategy. This leads to discuss the work by Rui Han on AdaptivConfig\cite{han2019adaptiveconfig}  which successfully reduced the job latencies compared to different static configurations.They have also highlighted the need for dynamic configurations to handle large cloud workloads. We are moving towards an era where efficient energy consumption will be key to sustainable future growth. But we are still using some traditional scheduling policies, aim of this paper is to understand how the same policies are relevant with the sustainable needs of future generations.

\section{Methods}
Scheduling is done to execute threads on the CPU fairly in computing systems. Under ideal conditions, any thread that is in the waiting state should not use CPU time. The effectiveness of work completion can be facilitated by the application of two suitable scheduling rules. The main policies used by Ubuntu are FIFO and RR. First in, first out (FIFO) is an interface designed for real-time applications, where the initial request process will be handled in order. Threads that are not time-sliced run until blocked. Threads of higher priority always prevail over those with lower priority. When a thread gets blocked, it is moved to the end of the priority list. Round robin is a variation of FIFO that adds time-slicing. After a certain period, threads are preempted and placed at the back of the priority queue. Once resumed, they finish their remaining time.

The deadline scheduling function in the Linux 3.14 kernel is based on the Global Earliest Deadline First algorithm \cite{singh2010algorithm}. In this specific scheduling method, threads are characterized by parameters called runtime, deadline, and period. The scheduling condition that must be satisfied is $\text{runtime} \leq \text{schedule deadline} \leq \text{schedule period}$. This type also uses throttling to counterbalance resource overuse. It is the default time-sharing scheduler for Linux, with scheduled threads determined by their nice value. The system tries to allocate CPU time fairly across threads. The range of nice values is from -20 to +19, and the impact of these variables is most noticeable on more recent kernel versions. Idle scheduling is another concept for background jobs, focusing on tasks with lower priority. A Linux Ubuntu system is used to understand the differences between round-robin (RR) scheduling and first-in, first-out (FIFO) scheduling for real-time processes. These observations are made using computationally expensive programs, with results visualized in graphs. Each line in the figure shows the scheduling results of clustering processes at different time points, along with their performance based on PowerTop report data.

    \begin{figure}[h]
        \centering
        \includegraphics[scale=0.3]{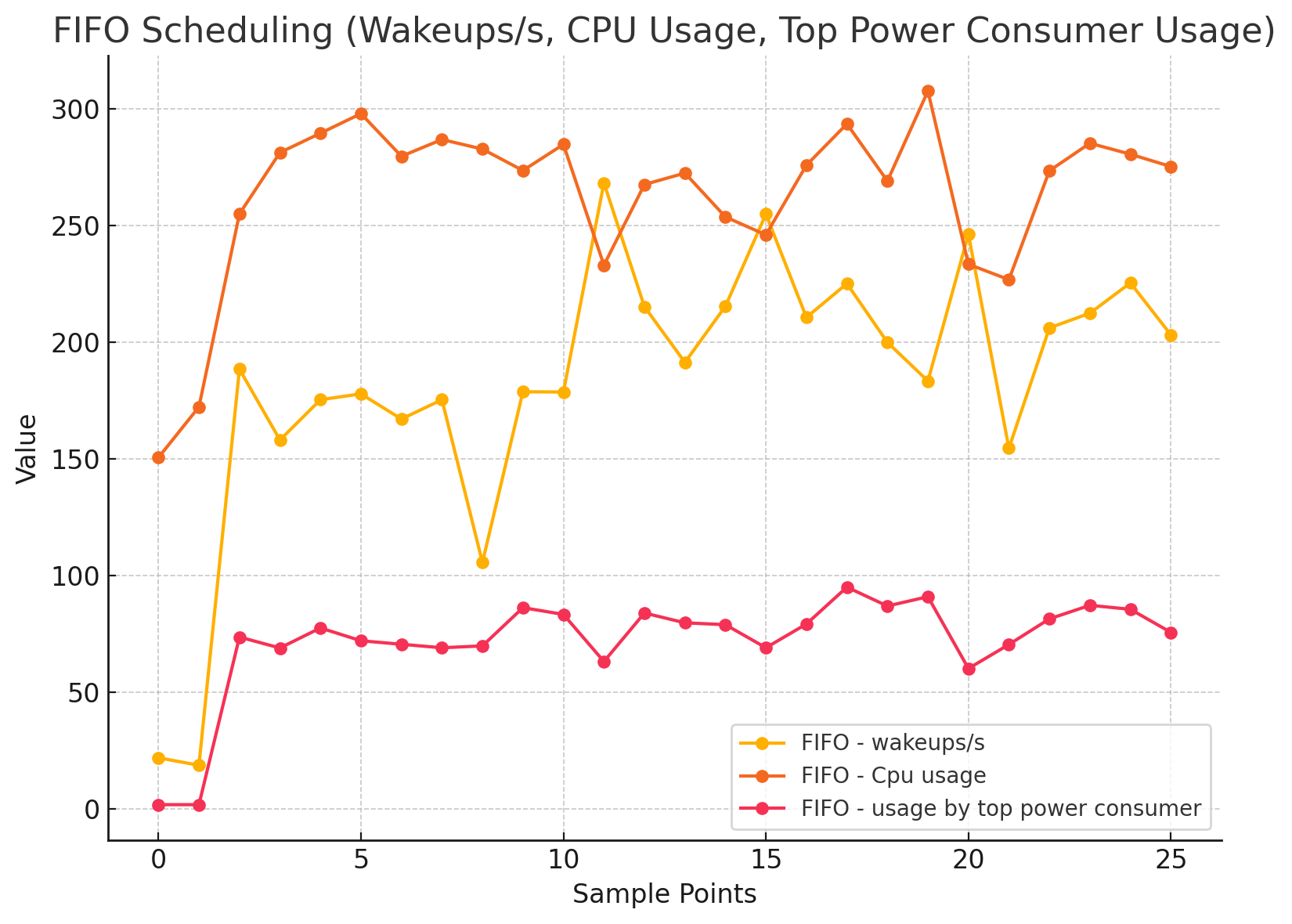}  
        \caption{Image 1}
        \label{fig:output1}
    
        \includegraphics[scale=0.3]{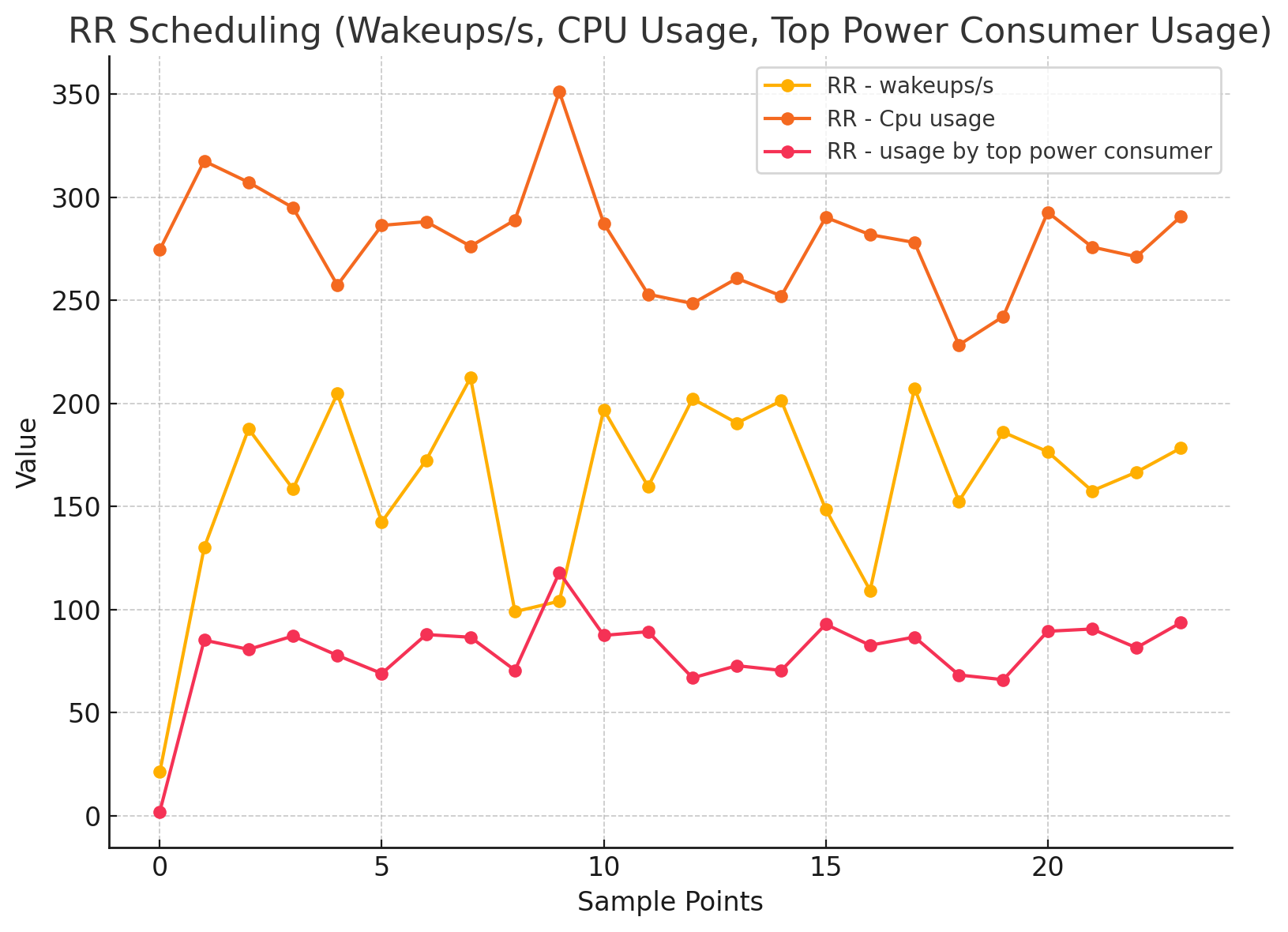}  
        \caption{Image 2}
        \label{fig:output2}
    
        \includegraphics[scale=0.3]{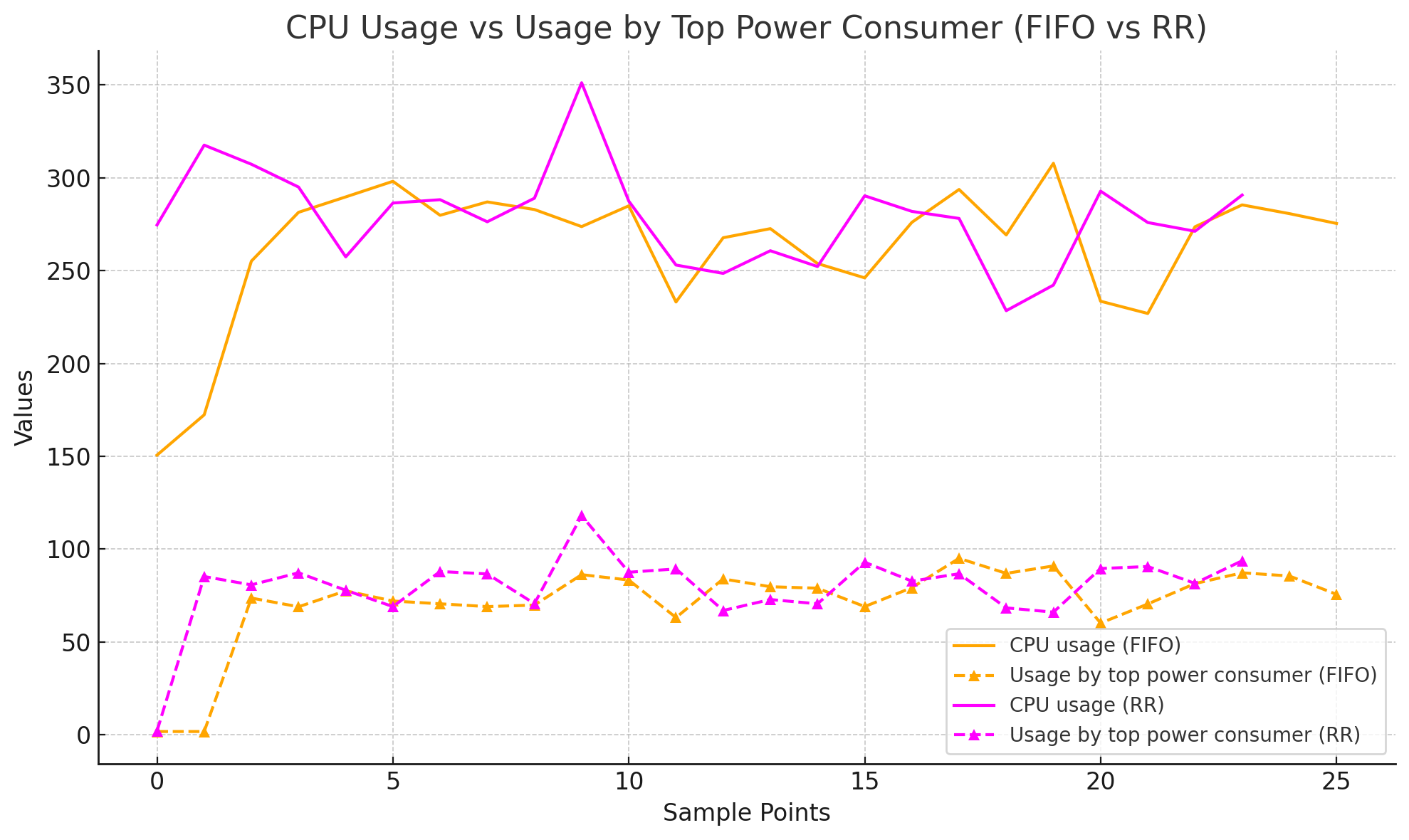}  
        \caption{Image 3}
        \label{fig:output3}
    
        \includegraphics[scale=0.3]{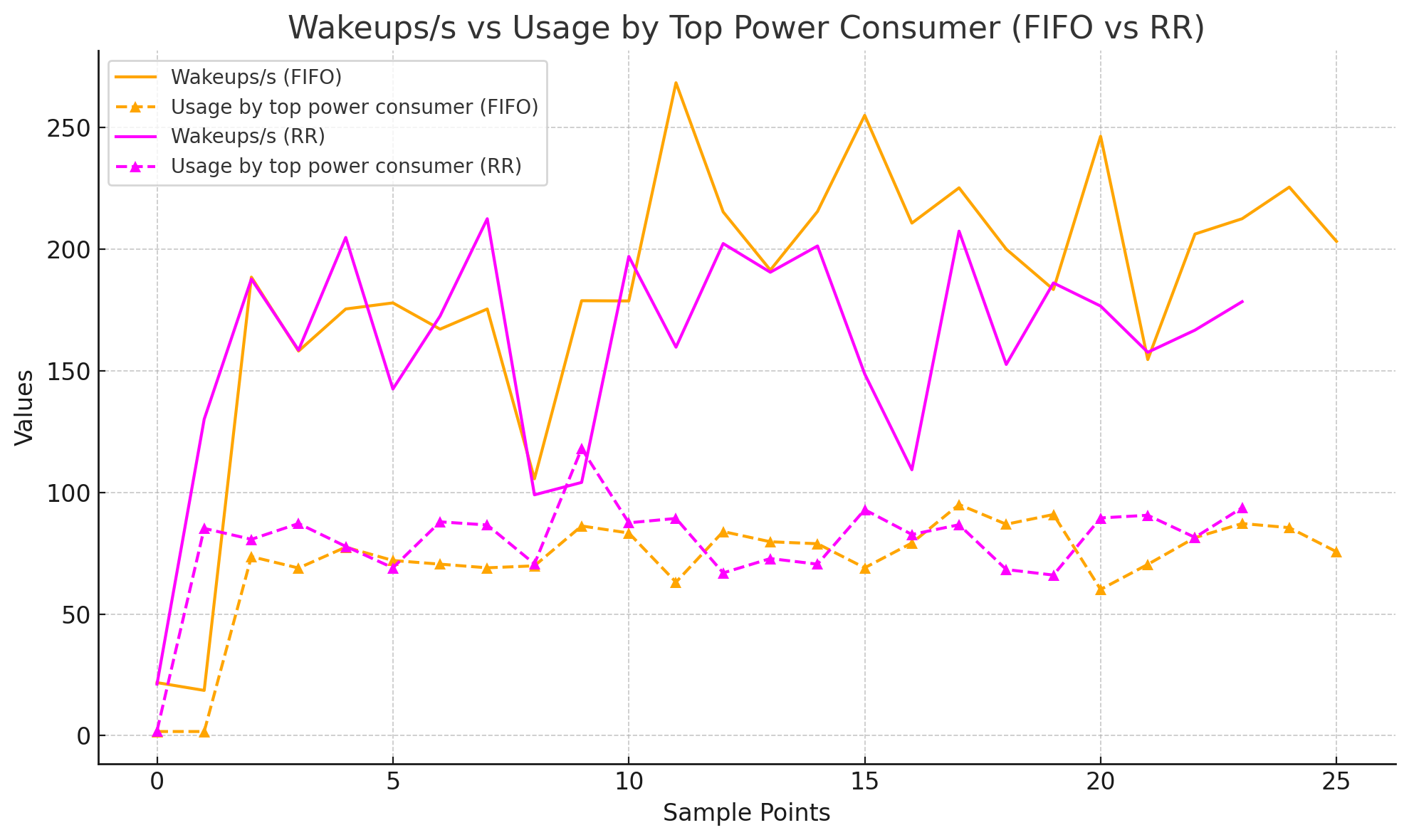}  
        \caption{Image 4}
        \label{fig:output4}
        
    \end{figure}

    \begin{figure}[h]
        \centering
        \includegraphics[scale=0.3]{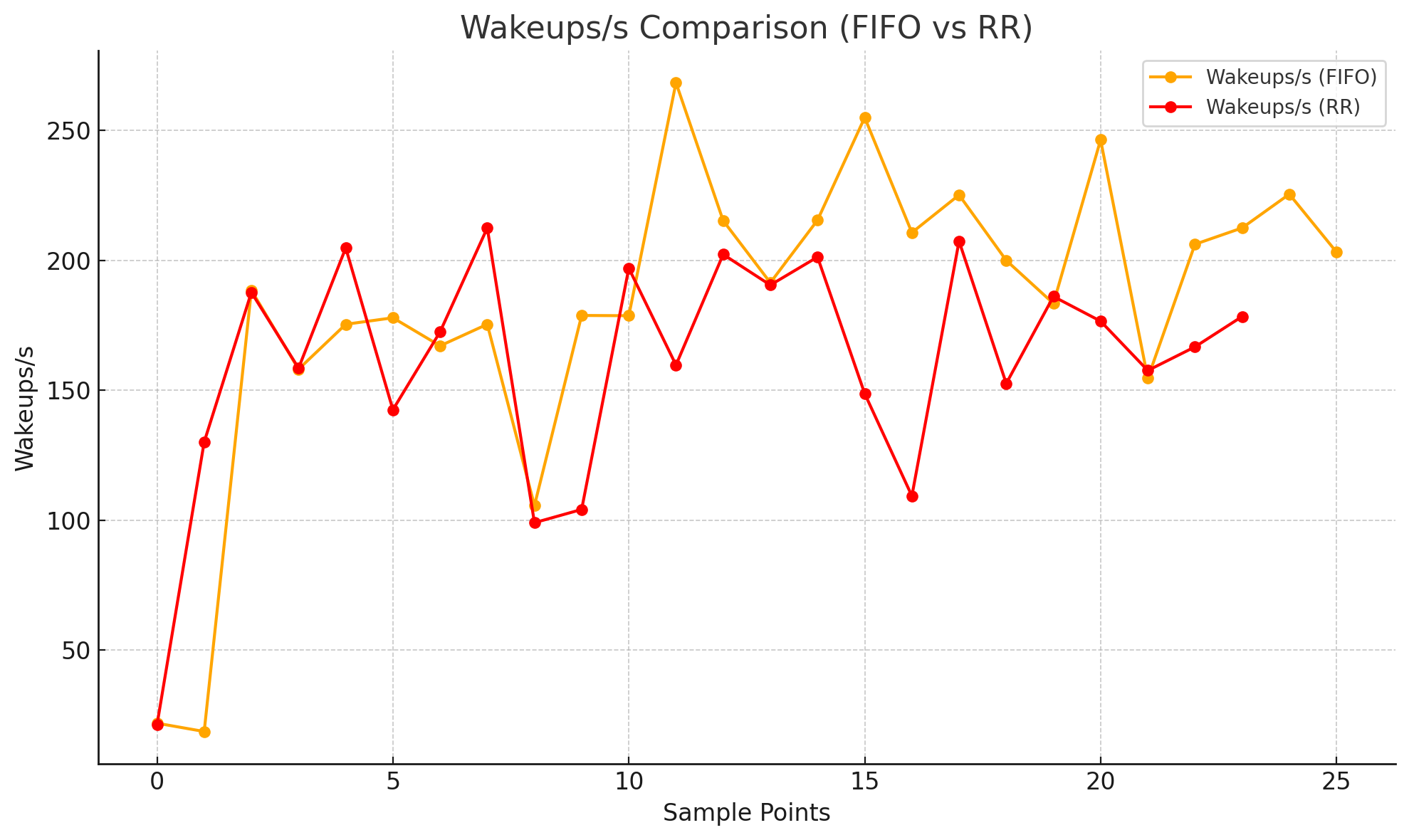}  
        \caption{Image 5}
        \label{fig:output1}
    
        \includegraphics[scale=0.3]{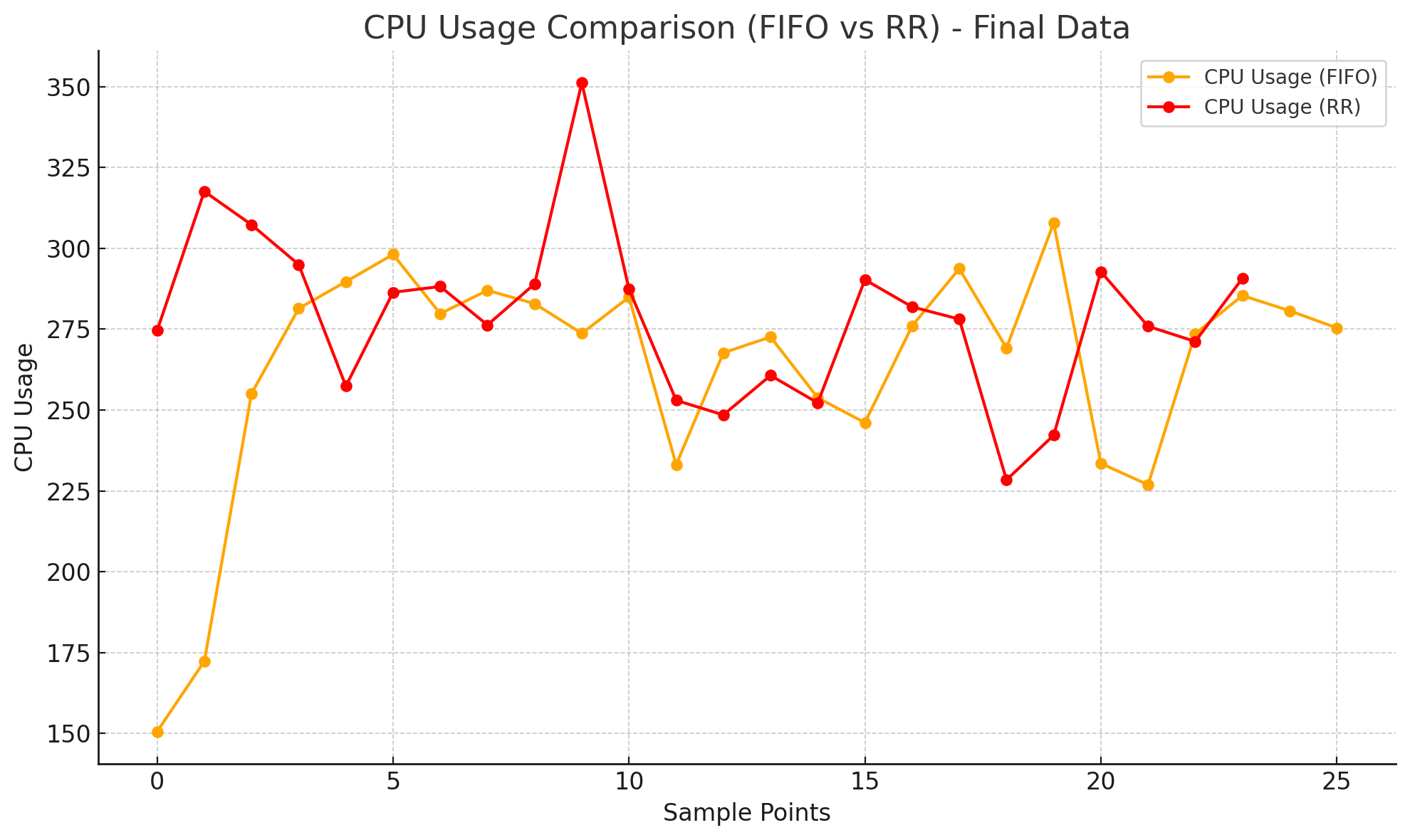}  
        \caption{Image 6}
        \label{fig:output2}
    
        \includegraphics[scale=0.3]{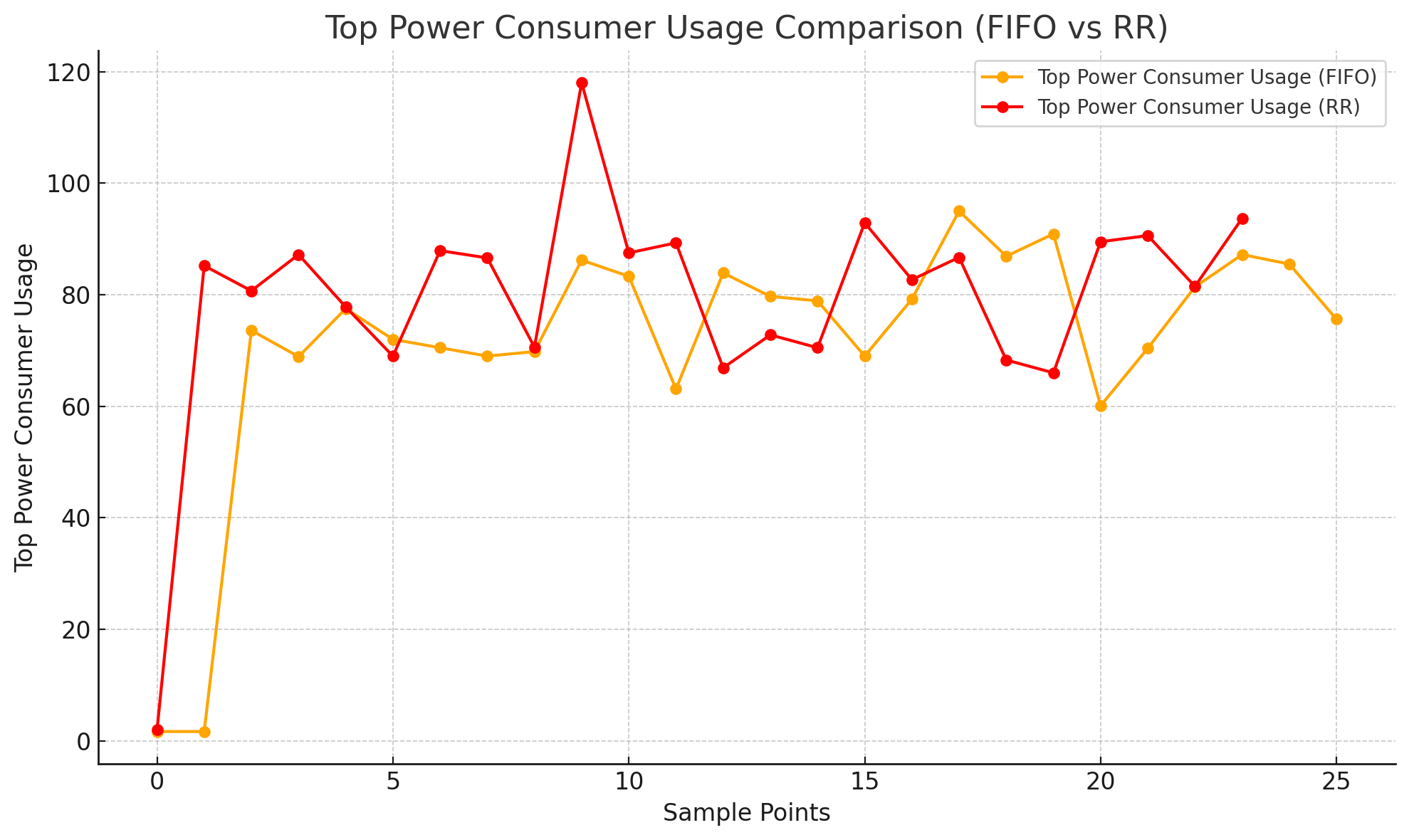}  
        \caption{Image 7}
        \label{fig:output3}
    
        \includegraphics[scale=0.3]{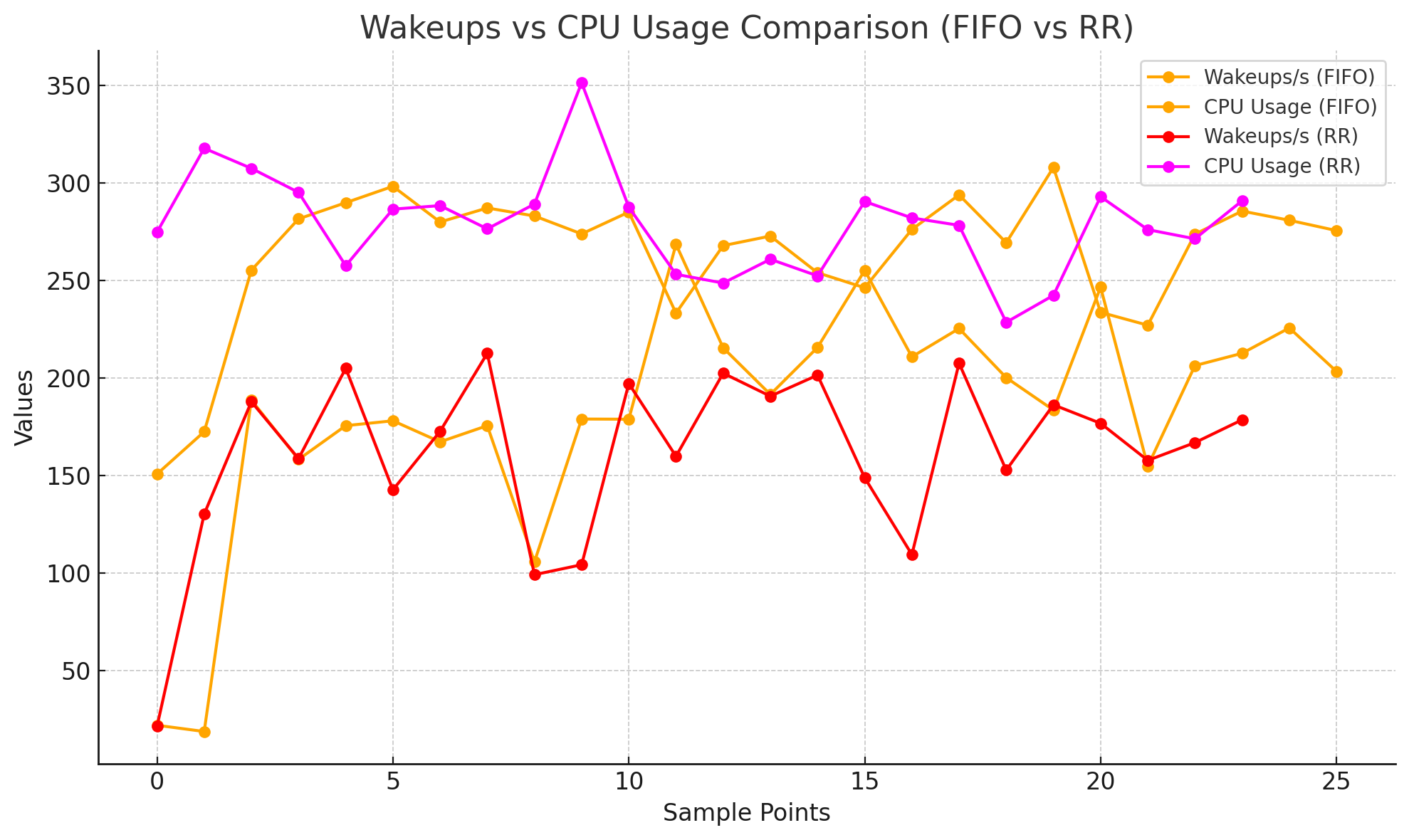}  
        \caption{Image 8}
        \label{fig:output4}
        
    \end{figure}

     \begin{figure}[h]
        \centering
        \includegraphics[scale=0.3]{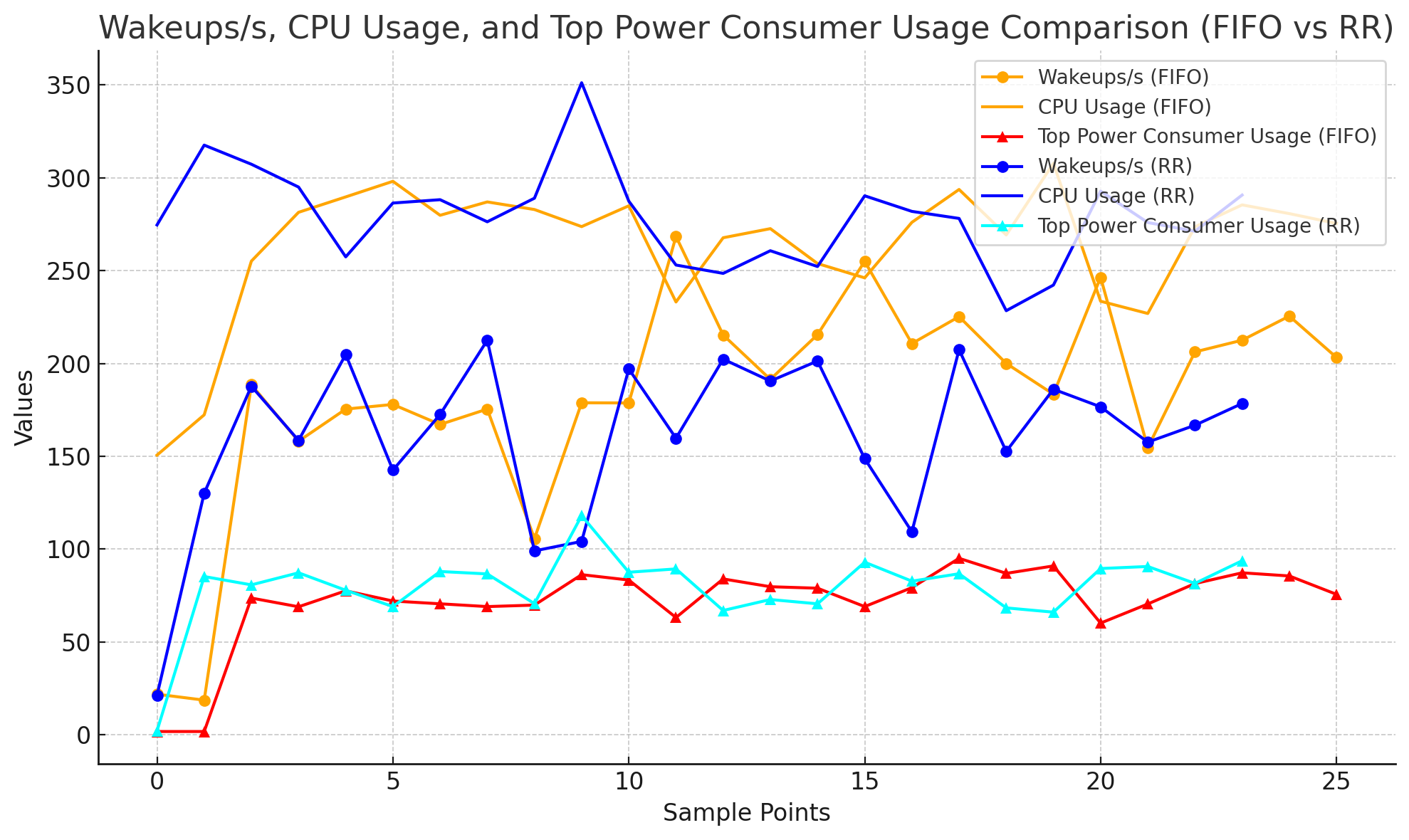}  
        \caption{Image 9}
        \label{fig:output1}
        \end{figure}

\section{Results}
The primary aim of comparing multiple process scheduling algorithms is to ascertain their efficiency, namely how soon the scheduler can ensure the completion of activities. Conceptually, the First-In, First-Out (FIFO) method should be the most efficient, however there are other variables that influence the efficiency of task execution. The results primarily compare the FIFO and RR scheduling policies of the presented methods. There are around 100 analysis of powertop findings from a Linux system, comparing the performance of FIFO and RR algorithms. These measurements are collected at random intervals with primarily processes operating as read-intensive and write-intensive data pipelines using Spark. It also involves training two machine learning models, namely CNN\cite{chua1997cnn} and LSTM \cite{sundermeyer2012lstm}, rendered into operating processes. These are computationally intensive generic procedures that may accurately compare scheduling algorithms based on statistical analysis to determine which one is more efficient. An initial conclusion that may be made contradicts the commonly held notion that convolution neural network (CNN) \cite{chua1997cnn} models or any multi-level (ML) procedure cannot be executed efficiently on a multithreaded system. Although to some extent this statement is accurate, sequential computations in a multilayer model are not appropriate for dispersed environments. However, this limitation mostly arises from the irregular nature of model training. Emphasizing the training of machine learning models, there are aspects of the model that may be enhanced, such as the combination of reading and writing data. Implementing these measures can significantly impact the training duration for each model if there are several models with data sizes below 100 GBs. Tensorflow \cite{abadi2016tensorflow} facilitates particular distributed computing tasks by utilizing tf.distribute.This strategy involves the division of large datasets into batches for processing. In the context of large-scale systems, spark pipelines or data pipelines can be designed to spread both read and write operations among several multi-level switches (MUS). The topic at hand is whether process scheduling can assist with this procedure. The use of FIFO scheduling instead of RR scheduling may effectively decrease the scheduling overhead and enhance the efficiency of a system, whether it is a large-scale system or a small system with over-time. This may be elucidated by referring to the reading extracted from the Ubuntu system. This is derived from data obtained from Powertop while the processes were executed using FIFO and Round Robin scheduling methodology. The FIFO algorithm exhibited a greater average wakeups-per-cpu ratio in comparison to Round Robin. These findings suggest that FIFO encounters a higher number of wakeups in relation to CPU consumption compared to RR. Furthermore, the data indicates that FIFO has a higher frequency of wakeups per CPU consumption, particularly at lower CPU usage levels. Repeatedly, RR constantly demonstrates greater average CPU use in comparison to FIFO.This implies that RR uses a greater amount of CPU resources.The wakeups-per-cpu use ratio for both FIFO and RR indicates that FIFO exhibits a higher rate of handling wakeups in comparison to CPU consumption.This may indicate improved effectiveness in managing activities that need frequent awakenings, but the specific characteristics of the workload may differ.Frequent superfluous wakeups lead to the inefficient usage of CPU resources.Hence, considering computationally demanding tasks, it is more advantageous to employ FIFO rather than RR. Additional conclusion drawn from the data reveals that a larger FIFO ratio implies a system that wakes up frequently but consumes less CPU each wakeup, whereas RR entails more CPU workload.This may be demonstrated by the level of total processing power consumed by the resource-intensive job. The program with the highest CPU consumption is indicated at the top of the powertop measurements. The CPU consumption is 8 percent higher for RR compared to FIFO.The mean utilization for RR is 79.33, whereas for FIFO it is 71.58. Indeed, this verifies that RR is capable of achieving a higher productivity per wakeup in comparison to FIFO.
\section{Energy Considerations}
When we talk about energy considerations for the RR and FIFO datasets, we focus on wakeups. Frequent wakeups imply that the CPU must constantly resume from idle states to handle tasks which will consume more energy. RR is using more energy per wakeup.But the system is waking up less frequently for FIFO but when its waking up it consumes more energy due to heavier tasks. FIFO would consume more energy overall during active periods due to higher CPU but it might save energy during idle periods as wakeups are in frequent. FIFO wakes up the cpu more frequently but performs lighter tasks This is more energy efficient if the goal is to handle many small tasks without consuming too much power per task. However the frequent wakeups can hinder energy-saving measures,leading to more power consumption. 
\section{Conclusions and Future Work}
Based on the synchronization techniques used in this research, the results show that using RR is more energy efficient. This allows for increased productivity. Evolutionarily, FIFO can also improve energy efficiency. With this data structure, one can develop new schedulers that target only cloud clusters. In a cloud environment, cluster administrators care more about the distribution of resources than computational performance. In the future, a more nuanced investigation of the deadline scheduling policy will be added to this work. Future research should also explore the contribution of rules to cloud schedulers. However, there are challenges in quantifying the energy efficiency of cloud-native settings. One day, it might be possible to find ways to measure them and fit them into the existing framework.

\appendices
\section{Details on python code}
The scheduling policies were tested against python code and spark data-pipeline in a multi-threaded bash script. Python code used the training for the popular MNIST dataset and spark pipelines used electric vehicle data in USA. The pipeline is designed to transform the data to understand the distribution of vehicles in California. Baseline was established by running stress tests for various policies.


\section*{Acknowledgment}
Thank you Baeldung blogs \cite{baeldung} for writing detailed blogs on topics like Process Scheduling in Ubuntu and Linux kernel functions, it helped me greatly in understanding the schedulers used by Ubuntu. Some of the sentences in Introduction,Method and Conclusion have been grammatically and linguistically enhanced by ChatGpt.

\ifCLASSOPTIONcaptionsoff
  \newpage
\fi




\begin{thebibliography}{}
\providecommand{\url}[1]{#1}
\csname url@samestyle\endcsname
\providecommand{\newblock}{\relax}
\providecommand{\bibinfo}[2]{#2}
\providecommand{\BIBentrySTDinterwordspacing}{\spaceskip=0pt\relax}
\providecommand{\BIBentryALTinterwordstretchfactor}{4}
\providecommand{\BIBentryALTinterwordspacing}{\spaceskip=\fontdimen2\font plus
\BIBentryALTinterwordstretchfactor\fontdimen3\font minus
  \fontdimen4\font\relax}
\providecommand{\BIBforeignlanguage}[2]{{%
\expandafter\ifx\csname l@#1\endcsname\relax
\typeout{** WARNING: IEEEtran.bst: No hyphenation pattern has been}%
\typeout{** loaded for the language `#1'. Using the pattern for}%
\typeout{** the default language instead.}%
\else
\language=\csname l@#1\endcsname
\fi
#2}}
\providecommand{\BIBdecl}{\relax}
\BIBdecl

\end{thebibliography}


\begin{thebibliography}{1}

\bibitem{stallings2018operating}
W.~Stallings, \emph{Operating Systems: Internals and Design Principles}, Pearson, 2018.

\bibitem{akgun2014energy}
O.~T.~Akgun, D.~G.~Down, and R.~Righter, "Energy-aware scheduling on heterogeneous processors," \emph{IEEE Transactions on Automatic Control}, vol.~59, no.~3, pp.~599--613, 2014, doi: \href{https://doi.org/10.1109/TAC.2013.2286756}{10.1109/TAC.2013.2286756}.


\bibitem{tanenbaum2015modern}
A.~S.~Tanenbaum and H.~Bos, \emph{Modern Operating Systems}, Pearson, 2015.

\bibitem{verma2015large}
A.~Verma, L.~Pedrosa, M.~R.~Korupolu, D.~Oppenheimer, E.~Tune, and J.~Wilkes, "Large-scale cluster management at Google with Borg," in \emph{Proceedings of the Tenth European Conference on Computer Systems (EuroSys)}, 2015, pp. 1--17.

\bibitem{vavilapalli2013apache}
V.~K.~Vavilapalli \emph{et al.}, "Apache Hadoop YARN: Yet Another Resource Negotiator," in \emph{Proceedings of the 4th Annual Symposium on Cloud Computing (SoCC)}, 2013, article no. 5.

\bibitem{ousterhout2013sparrow}
K.~Ousterhout, P.~Wendell, M.~Zaharia, and I.~Stoica, "Sparrow: Distributed, low latency scheduling," in \emph{Proceedings of the Twenty-Fourth ACM Symposium on Operating Systems Principles (SOSP)}, 2013, pp. 69--84.

\bibitem{delgado2015mercury}
P.~Delgado and D.~Epema, "Mercury: Hybrid centralized and distributed scheduling in large shared clusters," in \emph{Proceedings of the 2015 IEEE 7th International Conference on Cloud Computing Technology and Science (CloudCom)}, 2015, pp. 1--10.

\bibitem{chen2018hawk}
H.~Chen, W.~Wang, and S.~Chen, "Hawk: Hybrid job and resource management for data analytics in the cloud," \emph{IEEE Transactions on Parallel and Distributed Systems}, vol.~29, no.~7, pp. 1489--1502, 2018.

\bibitem{delgado2016eagle}
P.~Delgado, F.~Dinu, D.~Didona, and W.~Zwaenepoel, "Eagle: A better hybrid data center scheduler," in \emph{Proceedings of the IEEE International Conference on Cloud Computing Technology and Science ({C}loud{C}om)}, 2016, pp.~1--10.

\bibitem{haidri2020cost}
R.~A.~Haidri, C.~P.~Katti, and P.~C.~Saxena, "Cost effective deadline aware scheduling strategy for workflow applications on virtual machines in cloud computing," \emph{Journal of King Saud University - Computer and Information Sciences}, vol.~32, no.~6, pp.~666--683, 2020. DOI: \href{https://doi.org/10.1016/j.jksuci.2017.10.009}{10.1016/j.jksuci.2017.10.009}.



\bibitem{delimitrou2013paragon}
C.~Delimitrou and C.~Kozyrakis, "Paragon: QoS-aware scheduling for heterogeneous datacenters," \emph{ACM SIGARCH Computer Architecture News}, vol.~41, no.~1, pp. 77--88, 2013.

\bibitem{delimitrou2014quasar}
C.~Delimitrou and C.~Kozyrakis, "Quasar: Resource-efficient and QoS-aware cluster management," in \emph{Proceedings of the Nineteenth International Conference on Architectural Support for Programming Languages and Operating Systems (ASPLOS)}, 2014, pp. 127--144.

\bibitem{zhao2015alsched}
H.~Zhao and W.~Wang, "AlSched: A heterogeneity-aware virtual machine scheduler for analytical workloads," in \emph{Proceedings of the 2015 IEEE International Conference on Big Data (Big Data)}, 2015, pp. 355--360.

\bibitem{merkel2006balancing}
A.~Merkel and F.~Bellosa, "Balancing power consumption in multiprocessor systems," in \emph{Proceedings of the 1st ACM SIGOPS/EuroSys European Conference on Computer Systems 2006}, 2006, pp. 403--414.

\bibitem{dhiman2007performance}
G.~Dhiman and T.~Rosing, "Performance and energy trade-offs for multitasking environments using temporal partitioning," in \emph{Proceedings of the 44th Annual Design Automation Conference (DAC)}, 2007, pp. 297--302.

\bibitem{singh2010algorithm}
S.~Singh and I.~Chana, "A survey of scheduling techniques in cloud computing environment," in \emph{Proceedings of the 2010 International Conference on Advances in Computing and Communications (ACC)}, 2010, pp.~394--403.

\bibitem{chua1997cnn}
L.~O.~Chua, "CNN: A paradigm for complexity," \emph{International Journal of Bifurcation and Chaos}, vol.~7, no.~10, pp.~2219--2425, 1997.

\bibitem{sundermeyer2012lstm}
M.~Sundermeyer, R.~Schl\"{u}ter, and H.~Ney, "LSTM neural networks for language modeling," in \emph{Proceedings of the Thirteenth Annual Conference of the International Speech Communication Association (Interspeech 2012)}, Portland, OR, USA, 2012, pp.~194--197.

\bibitem{abadi2016tensorflow}
M.~Abadi, P.~Barham, J.~Chen, \emph{et al.}, "TensorFlow: A system for large-scale machine learning," in \emph{Proceedings of the 12th USENIX Symposium on Operating Systems Design and Implementation (OSDI)}, Savannah, GA, USA, 2016, pp.~265--283.


\bibitem{chandio2019comparative}
A.~A.~Chandio \emph{et al.}, "A comparative study of energy-efficient virtual machine scheduling algorithms for high-performance computing environments," \emph{Sustainable Computing: Informatics and Systems}, vol.~24, p.~100352, 2019.

\bibitem{han2019adaptiveconfig}
R.~Han, X.~Chen, and W.~Li, "AdaptiveConfig: Workload-driven dynamic configuration tuning for big data applications," \emph{IEEE Transactions on Parallel and Distributed Systems}, vol.~30, no.~12, pp. 2880--2893, 2019.

\bibitem{zakrya2020comparative}
A.~Zakrya, F.~A.~Omara, and M.~A.~AbdelAzim, "A comparative study on global and partitioned scheduling algorithms for real-time multiprocessor systems," \emph{International Journal of Computer Applications}, vol.~176, no.~28, pp. 34--39, 2020.

\bibitem{baeldung}
E.~Paraschiv, "Baeldung: Learn Java, Spring, and More," \emph{Baeldung}, [Online]. Available: \url{https://www.baeldung.com}. [Accessed: September 19, 2024].



\end{thebibliography}
%

\begin{IEEEbiography}[{\includegraphics[width=1in,height=1.25in,clip,keepaspectratio]{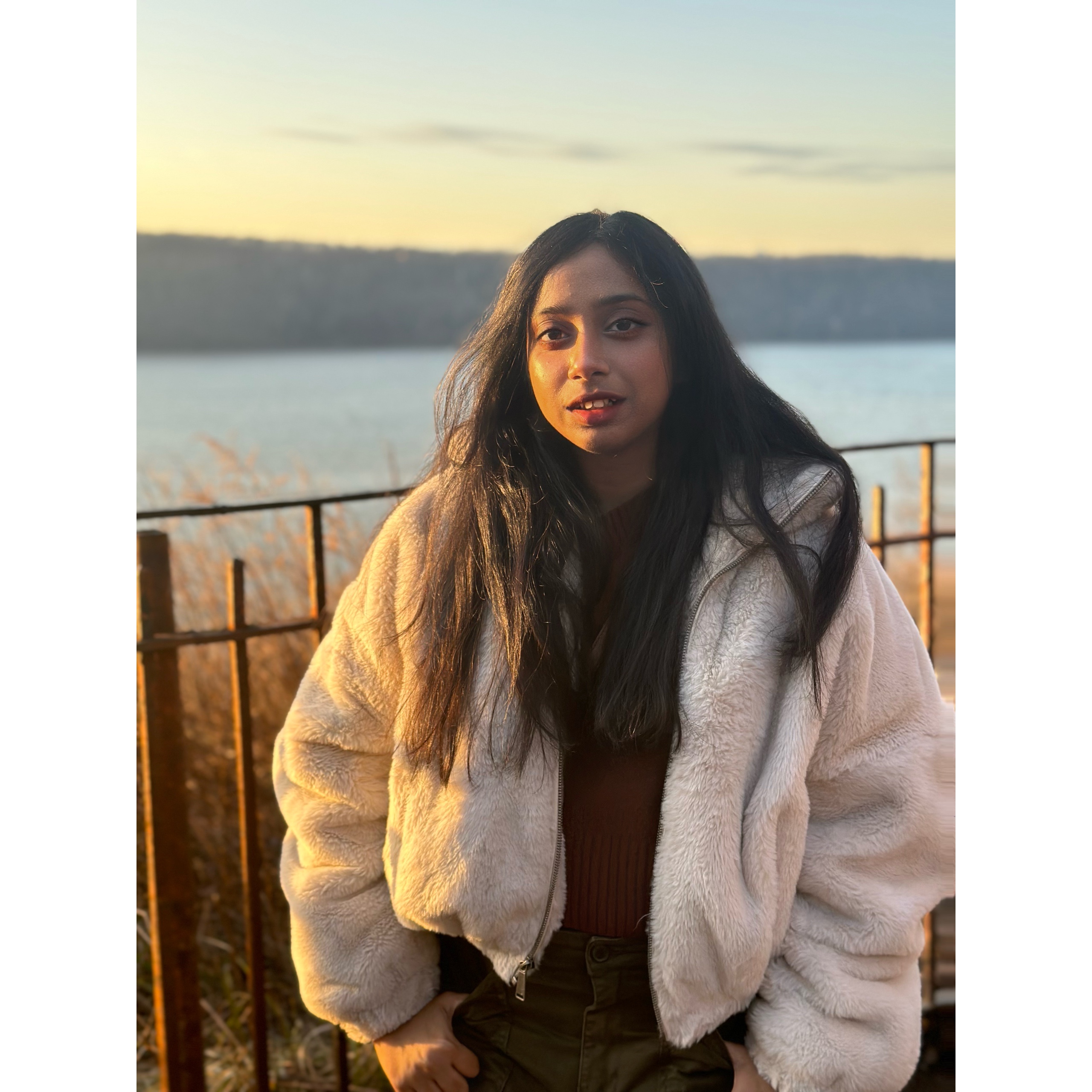}}]{Malobika Roy Choudhury}
Malobika Roy Choudhury is a researcher with a Master's degree from Arizona State University,  AZ 85281 USA. Her research interests include operating systems,ML workloads,algorithms and energy-efficient computing. She can be reached at \texttt{malobika.roychoudhury@gmail.com}.
\end{IEEEbiography}

\begin{IEEEbiography}{Akshat Mehrotra}
Akshat Mehrotra is an engineering manager pursuing a Master's degree from Harrisburg University, PA 039483 USA. His expertise lies in system optimization and sustainable lean designs. Contact him at \texttt{akshat08121989@gmail.com}.
\end{IEEEbiography}





\end{document}